\documentclass[jkps,preprint,fleqn]{revtex4} 
\usepackage{graphicx}
\usepackage{amssymb}
\usepackage{amsmath}
\usepackage{bm}

\usepackage{kotex}

\pdfoutput=1

\usepackage{float} \usepackage{graphicx}
\usepackage{epstopdf}
\usepackage{graphics}
\usepackage{caption}
\usepackage{subfigure}
\usepackage{epsfig}
\usepackage{color}
\usepackage{multirow,array}

\usepackage{tensor}

\usepackage{adjustbox}  

\usepackage{tabularray}

\usepackage{hyperref}
\hypersetup{
	colorlinks=true,
	linkcolor=blue,
	filecolor=magneta,      
	urlcolor=blue,
}

\begin{document}
\setcounter{page}{0}
\title[]{The significance of measuring cosmological time dilation in the Dark Energy Survey Supernova Program }
\author{Seokcheon \surname{Lee}}
\email{skylee@skku.edu}
\affiliation{Department of Physics, Institute of Basic Science, Sungkyunkwan University, Suwon 16419, Korea}

\date[]{Received }

\begin{abstract}
In the context of the dispersion relation $c = \lambda \nu$ and considering an expanding universe where the observed wavelength today is redshifted from the emitted wavelength by $\lambda_{0} = \lambda_{\text{emit}} (1+z)$, to keep $c$ constant, it must be that  $\nu_{0} = \nu_{\text{emit}} /(1+z)$. However, although the theory for wavelength in the RW metric includes the cosmological redshift, the same is not simply deduced for frequency (the inverse of time). Instead, cosmological time dilation $T_{0} = T_{\text{emit}} (1+z)$ is an additional assumption made to uphold the hypothesis of constant speed of light rather than a relation directly derived from the RW metric. Therefore, verifying cosmological time dilation observationally is crucial. The most recent data employing supernovae for this purpose was released recently by the Dark Energy Survey. Results from the i-band specifically support variations in the speed of light within 1-$\sigma$. We used these observations to investigate variations in various physical quantities, including $c$ and $G$, using the minimally extended varying speed of light model. The speed of light was $0.4$\% to $2.2$\% slower, and Newton's constant may have decreased by $1.7$\% to $8.4$\% compared to their current values at redshift $2$. These findings, consistent with previous studies, hint at resolving tensions between different $\Lambda$CDM cosmological backgrounds but are not yet conclusive evidence of a varying speed of light, as the full-band data aligns with standard model cosmology.  However, the data remains valuable for testing variations in fundamental constants over cosmic time. Future analyses, particularly with more refined redshift data, may provide clearer insights into these potential changes.
\end{abstract}



\maketitle


\section{Introduction}\label{sec1}

The Robertson-Walker (RW) metric, derived from the cosmological principle (CP) and Weyl's postulate, characterizes the $\Lambda$CDM cosmological model \cite{Islam01,Narlikar02,Hobson06,Roos15}. This model theoretically predicts that dimensionful quantities such as wavelength and temperature undergo cosmological redshift due to cosmic expansion. Traditionally, the speed of light is assumed constant in this framework, from which the relation for cosmological time dilation (CTD) is derived. However, even if we adopt CP and Weyl's postulate, they do not specify the relation for CTD within the RW metric. Therefore, if observations find CTD deviating from the standard model cosmology (SMC) predictions, there is a possibility that the speed of light may vary with cosmic time \cite{Lee:2020zts,Lee:2022heb,Lee:2023bjz,Lee:2024part}.

 Using the dispersion relation $c = \lambda \nu$ within the framework of cosmic expansion, where $\lambda_{0} = \lambda_{\text{emit}} (1+z)$, maintaining the constancy of the speed of light $c$ requires that $\nu_{0} = \nu_{\text{emit}} (1+z)^{-1}$. This relationship ensures that the wavelength $\lambda$ stretches proportionally with the scale factor $a$ (i.e., $(1+z)^{-1}$) as the universe expands \cite{Weinberg:2008}. However, while the theory of wavelength evolution under the RW metric naturally includes cosmological redshift, the situation for frequency, which is the inverse of time, is more complex. Without additional assumptions about the variation of the magnitude of the speed of light, the cosmological relationship for frequency cannot be specified solely from the RW metric. Unlike the straightforward deduction for wavelength, the assertion of CTD $T_{0} = T_{\text{emit}} (1+z)$ is an assumption rather than a direct consequence derived from the RW metric. Based on the SMC, it is predicted that the light curves of sources with intrinsically constant periodicity will stretch by $(1 + z)$ along the time axis. Therefore, empirically verifying whether CTD exactly follows this relationship is crucial for validating the broader implications of the theoretical constructs of modern cosmology.

Further investigation, however, presents the intriguing idea that $T_{0} = T_{\text{emit}} (1+z)^{1+\beta}$ could be the variation of the observed frequency $\nu_{0}$. This variation suggests that in an expanding universe, the speed of light $c$ observed today, $c_{0}$, could differ from its emitted counterpart $c_{\text{emit}}$ by a factor $(1+z)^{\beta}$, where $\beta$ represents a potential deviation from zero. We have proposed a model called the minimally extended varying speed of light (meVSL), and the parameter of this model is $b = -4\beta$~\cite{Lee:2020zts,Lee:2022heb,Lee:2023bjz,Lee:2024part}.

If $\beta \neq 0$, $c$ is not constant over cosmological time scales. Alternatively, $c$ might change as the universe expands, indicating a dynamic interaction between the scale factor $a$, wavelength $\lambda$, and frequency $\nu$. Such a scenario challenges the traditional view of $c$ as an immutable constant and underscores the importance of investigating CTD, where $T_{0} = T_{\text{emit}} (1+z)^{1+\beta} $, in observational cosmology.

The core of the argument is that the lack of a theory for CTD in an expanding universe implies that the speed of light could vary, increasing or decreasing like the wavelength, rather than being a constant, indicating that this is not just an issue of units.

There have been several ambitious projects aimed at measuring CTD, which is a fundamental prediction of an expanding universe. Direct observation of TD involves analyzing the decay time of distant type Ia supernovae (SNeIa) light curves (LCs) and their spectra, which have been the focus of numerous studies \cite{Leibundgut:1996qm,SupernovaSearchTeam:1997gem,Foley:2005qu,Blondin:2007ua,Blondin:2008mz,DES:2024vgg}. An alternative method of determining CTD is to look at how peak-to-peak timescales in gamma-ray bursts (GRBs) stretch; these observations offer important information on how these high-energy events behave over large cosmological distances \cite{Norris:1993hda,Wijers:1994qf,Band:1994ee,Meszaros:1995gj,Lee:1996zu,Chang:2001fy,Crawford:2009be,Zhang:2013yna,Singh:2021jgr}. Additionally, researchers have investigated the TD effect in the light curves of quasars (QSOs) situated at cosmological distances, attempting to detect the anticipated dilation \cite{Hawkins:2001be,Dai:2012wp,Lewis:2023jab}.

In this regard, the latest Dark Energy Survey (DES) supernova (SNa) program data was made public \cite{DES:2024vgg}. They employ two techniques to match the form of TD. In the first method, they minimized the flux scatter in stacked LCs to measure CTD. They divided all the SNa LCs' time axes by $(1 + z)^n$ and found that TD was present when a power of $n \approx 1$ minimized the scatter. They fit each SNa LC to a reference LC created by stacking supernovae (SNe) with matching observed bandpasses in the second method. They measured the observed LC width $w$ relative to this reference and found $n = 1.003 \pm 0.005 (\textrm{stat}) \pm 0.010 (\textrm{sys})$, further confirming TD and ruling out non-time-dilating models.

In the following section~\ref{sec:methods}, we introduce the latest CTD data obtained from SNeIa, GRBs, and QSOs. Additionally, we provide a brief explanation of the meVSL model. In Section~\ref{sec:results}, we specifically investigate the cosmic time evolution of the speed of light ($c$) and the gravitational constant ($G$) within the meVSL model using the CTD values obtained from the i-band of the DES~\cite{DES:2024vgg}. The final section~\ref{sec:Conc} discusses how the differences in redshift interpretation within the meVSL model compared to traditional methods can affect the interpretation of CTD data, along with future work and conclusions.

\section{Methods}
\label{sec:methods}

Accurate measurements of CTD are essential for two reasons: first, they provide proof of the expanding universe; second, they show whether the speed of light varies or stays constant in an expanding universe. Additionally, it can help deduce changes in other physical constants over cosmic time \cite{Lee:2020zts,Lee:2022heb,Lee:2023bjz,Lee:2024part}. Therefore, even if CTD provides little information on cosmological parameters within the context of the standard model cosmology (SMC), it is a critically significant observational quantity by itself. Of course, this requires an event or fluctuation with a known rest-frame duration observable at high redshift with sufficient accuracy to detect the predicted stretching by a factor of $(1 + z)^{1+\beta}$. Specifically, within the RW metric, whether the speed of light changes with the universe's expansion depends on how much CTD deviates from $(1+z)$, making precise measurement of this phenomenon extremely important~\cite{Lee:2024part}.

\subsection{Cosmological time dilation from standard clocks}
\label{subsec:CTDSC}

Astronomers identify them using LCs and SNeIa spectra, which show that they are white dwarf thermonuclear explosions in binary systems. By plotting apparent brightness versus time and analyzing LC widths, astronomers can determine the intrinsic behavior of SNeIa without TD effects. However, observed LCs are stretched due to TD, with an expected TD factor of $(1 + z)^{1+\beta}$, making the SNe appear to take longer to reach peak brightness and fade away. A large sample of SNeIa at various redshifts is studied to validate these observations. They observe through statistical fits the power law that minimizes dispersion in standardized LCs across different redshifts and supports the TD effect \cite{DES:2024vgg}.

The DES assessed SNeIa LCs detected in the $g$, $r$, $i$, and $z$ bands using the reference-scaling method. In the $i$-band, 1465 SNe passed quality cuts with a reduced chi-square value of $0.788$, fitting the TD as $(1+z)^n$ with $n = 0.988 \pm 0.008$. Across all bands, the relationship $(1 + z)^{1.003\pm0.005}$ for TD was derived from $1504$ unique SNe, with a reduced chi-square of $1.441$. It supports the TD effect with consistent results across different observed bands. These findings are summarized in Table~\ref{tab:CTDs}.

A standard clock with a measured timescale is necessary to look for CTD. However, in the case of other cosmic objects like quasars and GRBs, the complexity of the physical processes causing their variability makes it difficult to use them as standard clocks. Nevertheless, the standard clock method using QSos and GRBs has been consistently observed.

Measuring CTD from GRBs involves observing these intense cosmic explosions and analyzing their LCs. GRBs are classified into long bursts ($T_{90} > 2$ seconds) associated with core-collapse SNe and short bursts ($T_{90} < 2$ seconds) linked to mergers of compact objects. Space-based observatories like Swift and Fermi detect GRBs, and the redshift $z$ is determined by observing the afterglow in lower-energy bands and identifying spectral lines. The CTD effect predicts that the observed duration of GRBs should be stretched by a factor of $(1 + z)$. Researchers compare observed durations $T_{90}$ or $T_{50}$ against $(1 + z)^{1+\beta}$ to test for this stretching, using statistical methods to minimize uncertainties, such as stacking LCs in redshift bins and employing Bayesian analysis.

In analyzing $247$ GRBs, intrinsic scatter greater than $100$\% complicates establishing a clear relationship between $T_{50}/T_{90}$ and redshift  \cite{Singh:2021jgr}. The power-law index $\beta$ differs from the expected value of one within $1$-$\sigma$. By focusing on long GRBs and using geometric means to reduce outlier effects, a tighter scatter is observed. For $T_{50}$, the intrinsic scatter is less than $60$\%, and for $T_{90}$, it is consistent with $27$\% within a $68$\% confidence level, indicating the best-fit scatter value exceeds $100$\%. The power-law exponent $\beta$ aligns with a cosmological signature within 1-$\sigma$ for both $T_{50}$ and $T_{90}$. This analysis supports the presence of CTD, with the geometric mean providing the most consistent results ($n \equiv 1+\beta = 1.18^{+0.26}_{-0.36}$).

For QSOs, variability arises from stochastic processes in the relativistic disk around a supermassive black hole, influenced by factors like black hole mass, accretion rate, and observation wavelength. Measuring CTD involves compiling LCs showing how a QSO's brightness changes over time and determining the redshift $z$ to understand cosmic expansion. According to CTD, the observed timescales $\Delta t_{0}$ should be stretched by a factor of $(1 + z)^{1+\beta}$ compared to the rest-frame timescales $\Delta t_{\text{rest}}$ \cite{Lewis:2023jab}. They analyze LCs to identify variability patterns, considering systematic effects like observational biases and data quality. Using Diffusive Nested Sampling (DNest4), they found the posterior probability distribution for the hypothesis that quasar variability follows $(1 + z)^n$, discovering $n = 1.28^{+0.28}_{-0.29}$, consistent with the expected CTD.

 \begin{table}[h!]
 	\begin{center}
		\begin{tabular}{|c|c|c|c|} 
			\hline
			obs  & $1+\beta$ & $\#$ of samples & ref \\ \hline	
			\multirow{2}{*}{SNeIa} \quad i-band  & $0.988\pm0.008$  & 1465 & \multirow{2}{*}{\cite{DES:2024vgg}} \\ 
			\qquad \quad \,\, $4$-bands  & $1.003\pm0.005$ & 1504 &  \\ \hline
			\multirow{4}{*}{GRBs} \multirow{2}{*}{unbinned} \,\, $T_{50}$& $0.66^{+0.17}_{-0.17}$ & \multirow{4}{*}{$247$} &  \multirow{4}{*}{\cite{Singh:2021jgr}}  \\
				\hspace{3.0cm} $T_{90}$ & $0.52^{+0.15}_{-0.16}$ & & \\ 
			\qquad \quad	\multirow{2}{*}{binned} \quad \, $T_{50}$ &$1.18^{+0.26}_{-0.36}$  & & \\ 
				\hspace{3.0cm} $T_{90}$  & $0.97^{+0.29}_{-0.30}$ & & \\ \hline	
			QSOs & $1.28^{+0.28}_{-0.29}$ &190 & \cite{Lewis:2023jab} \\ \hline			
\end{tabular}
\end{center}
\caption{This table summarizes the most recent CTD data obtained from SNeIa, GRBs, and QSOs.}
\label{tab:CTDs}
 \end{table}

\subsection{The minimally extended varying speed of light model}
\label{subsec:meVSL}

Although general relativity (GR) is a theory concerning gravity, it is a local theory. Therefore, fundamentally, there are inherent difficulties in using GR to study the universe over time. One can solve this challenge with the RW metric \cite{Islam01,Narlikar02,Hobson06,Roos15}. This model starts from the postulate known as the CP, which states that the 3-dimensional space of the universe is homogeneous and isotropic at every moment. From this, one derives an equation satisfying the universe's spacetime at every moment $t_l$
\begin{align}
ds_{l}^2 = - c_l^2 dt^2 + a_l^2 \left[ \frac{dr^2}{1-Kr^2} + r^2  \left( d \theta^2 + \sin^2 \theta d \phi^2 \right)  \right] \quad \textrm{at} \,\,  t= t_l \label{dststar} \,,
\end{align}
where the scale factor $a_l$ and the speed of light $c_l$ in Eq.~\eqref{dststar} must remain constant to maintain homogeneity at a specific time $t_l$. One then introduces Weyl's postulate, which states that matter in the universe flows along geodesics. This concept helps define a universal time (\textit{i.e.}, cosmic time) for all observers in the universe. Essentially, this means that the paths of particles can be described as being at rest relative to this cosmic time, indicating that matter forms a consistent flow through space. Mathematically, Weyl's postulate suggests that the universe is a three-dimensional hypersurface evolving within the framework of GR. From this, we can expand equation~\eqref{dststar} as 
\begin{align}
ds^2 = - c(t)^2 dt^2 + a(t)^2 \left[ \frac{dr^2}{1-Kr^2} + r^2  \left( d \theta^2 + \sin^2 \theta d \phi^2 \right)  \right] \equiv - c(t)^2 dt^2 + a(t)^2 dl_{3\textrm{D}}^2 \label{dstgen} \,. 
\end{align} 
We write the speed of light as a function of time in equation~\eqref{dstgen}, which departs from the standard RW metric. This equation could look illogical or inaccurate at first. However, to consider the speed of light as a constant within the RW metric, an additional assumption about CTD is necessary, in addition to the CP and Weyl's postulate \cite{Lee:2020zts,Lee:2023bjz,Lee:2024part}. Thus, as shown below, the RW metric inherently allows for the possibility of the speed of light varying with time depending on the CTD relation. 

One obtains the cosmological redshift by using the geodesic equation for a light wave expressed as $ds^2 = 0$ in Eq. ~\eqref{dstgen}. $dl_{3\textrm{D}}$ is consistent over time because comoving coordinates are the same. Building on this foundation, we obtain the expression for radial light signals as
\begin{align}
d l_{3\textrm{D}} &= \frac{c(t_i) dt_i}{a(t_i)} \quad : \quad \frac{c_1 dt_1}{a_1} = \frac{c_2 dt_2}{a_2} \Rightarrow \begin{cases} c_1 = c_2 = c & \textrm{if} \quad \frac{dt_1}{a_1} = \frac{dt_2}{a_2} \qquad \textrm{SMC} \\ 
 c_1 = \frac{f(a_2)}{f(a_1)} \frac{a_1}{a_2} c_2 & \textrm{if} \quad \frac{dt_1}{f(a_1)} = \frac{dt_2}{f(a_2)} \quad \textrm{VSL} \\ c_1 = \left( \frac{a_1}{a_2}\right)^{\frac{b}{4}} c_2 & \textrm{if} \quad \frac{dt_1}{a_1^{1-\frac{b}{4}}} = \frac{dt_2}{a_2^{1-\frac{b}{4}}} \quad \textrm{meVSL}  \end{cases} \,, \label{dl3D}
\end{align}
where $f(a_i)$ specifies an arbitrary function of $a(t_i)$, and $d t_i = 1/\nu(t_i)$ represents the time interval between successive crests of light at $t_i$ (\textit{i.e.}, the inverse of the frequency $\nu_i$ at $t_i$). Thus, the RW metric implies that the speed of light can redshift like a wavelength depending on the form of CTD. A further supposition in the standard model cosmology (SMC) is the constant speed of light $c$. This assumption stems from SMC's reliance on GR, which treats $c$ as a constant. Therefore, the CTD between two hypersurfaces at $t = t_1$ and $t = t_2$ is directly related to the inverse of the scale factors $a(t)$ at those specific times as shown in Eq.~\eqref{dl3D}. However, this relationship is not based on any physical laws. In contrast, as hypothesized in this paper, if the speed of light varies with time, this relationship may no longer hold.

On the other hand, as one moves from one hypersurface to the next in an expanding universe, the scale factor rises, and other physical parameters, such as temperature and mass density, inevitably experience a cosmic redshift. However, in the RW metric, it is impossible to conclude about CTD based solely on the CP and Weyl’s postulate. Thus, it is necessary to incorporate the cosmic variable speed of light into the Einstein Field Equations (EFEs) and solve for solutions for this idea to form a coherent model. Our previous research has addressed such scenarios, particularly in the context of a model known as meVSL \cite{Lee:2020zts, Lee:2023bjz,Lee:2024part}. It also induces the cosmological evolution of other physical quantities and constants, including the Planck constant. Local thermodynamics, energy conservation, and other local physical phenomena, such as electromagnetic, are all included in the meVSL model. Permittivity, electric charge, and permeability cosmologically evolve throughout time according to the covariance of Maxwell equations in the meVSL model. These factors also contribute to the temporal evolution of various physical constants and quantities, as detailed in Table \ref{tab:LocalPhys}.

 \begin{table}[h!]
 	\begin{center}
		\begin{tabular}{|c||c|c|c|} 
	\hline
	local physics laws & Special Relativity & Electromagnetism & Thermodynamics \\
	\hline \hline
	quantities & $m = m_0 a^{-b/2}$ & $e = e_0 a^{-b/4}\,,  \lambda = \lambda_0 a\,, \nu = \nu_0 a^{-1+b/4}$ & $T = T_0 a^{-1}$ \\
	\hline
	constants & $c = c_0 a^{b/4}\,, G = G_0 a^{b}$ & $\epsilon = \epsilon_0 a^{-b/4}\,, \mu = \mu_0 a^{-b/4} $ & $k_{\textrm{B} 0}\,, \hbar = \hbar_0 a^{-b/4}$ \\
	\hline
	energies & $mc^2 = m_0 c_0^2$ & $h \nu = h_0 \nu_0 a^{-1}$ & $k_{\textrm{B}} T = k_{\textrm{B}} T_0 a^{-1}$ \\
	\hline
\end{tabular}
\end{center}
\caption{The cosmic evolutions of physical constants and values in the meVSL model are summarized below. These relations satisfy all known laws of local physics, including electromagnetic force, thermodynamics, and special relativity \cite{Lee:2020zts,Lee:2022heb,Lee:2023bjz,Lee:2024part}. In each quantity, the subscript $0$ denotes the present-day value of those quantities.}
\label{tab:LocalPhys}
\end{table}

\section{Results}
\label{sec:results}

The DES evaluated SNeIa LCs detected in the $g$, $r$, $i$, and $z$ bands using the reference-scaling method with precision to obtain CTD \cite{DES:2024vgg}. By analyzing data from all the bands combined, 1504 distinct SNeIa were included in the analysis. The analysis yielded a reduced chi-square value of $1.441$ and a time dilation relation of $(1 + z)^{1.003\pm0.005}$, which aligns with the SMC within the 1-sigma confidence level. However, this result only includes statistical errors. In analyzing supernova light curves, one might need to consider potential systematic effects that could influence the observed data. One such effect is the evolution of the stretch of supernova light curves as a function of redshift. After examining this factor, DES found that the impact is relatively small, with a likely upper limit of around $\sigma^{\textrm{sys}} = 0.01$. Therefore, we conduct the following analysis by dividing it into cases with and without this systematic error.

\subsection{all-band without systematic error}
\label{subsec:allbandnosys}

\subsubsection{$i$-band without systematic error}
\label{subsubsec:ibandnosys}

In contrast, 1465 SNe satisfied the quality criterion when the analysis was limited to the $i$ band. The decreased chi-square value for these SNeIa was $0.788$.  The DES analyzes $(1+z)^n$, with $n = 0.988 \pm 0.008$, provided the best match for the time dilation. This data suggests that the CTD may be affected by the VSL, as it deviates from the SMC within the $1$-$\sigma$ confidence interval. 

The implications of these findings suggest potential temporal variations in fundamental constants such as the speed of light and Newton's gravitational constant. These variations are critical as they may point towards new physics beyond the standard model. In the meVSL model, the CTD is given by 
\begin{align}
T_{0} = T (1+z)^{1-b/4} \label{T0meVSL} \,,
\end{align}
where the parameter $b = -4 \beta$. Thus, we can convert the result of the $i$ band of DES into the meVSL parameter $b$ as
\begin{align}
b = 0.048 \pm 0.032 \label{bmeVSL} \,.
\end{align}
We can use $c = c_0 (1+z)^{-b/4}$ and $G = G_0 (1+z)^{-b}$ to show the cosmological evolution of them. Figure 1 presents a thorough examination and a comparative depiction of these variations.  
\begin{align}
0.978 \leq \frac{c(z=2)}{c_0} \leq 0.996 \quad , \quad
0.916 \leq \frac{G(z=2)}{G_0} \leq 0.982 \label{cGz2} \,, 
\end{align}
These results are consistent with previous work suggesting that an effective Newton's constant $G_{\text{eff}}(z)$ decreasing with redshift may alleviate the observed tension between the Planck15 best-fit $\Lambda$CDM cosmological background and the $\Lambda$CDM background favored by growth $f \sigma_{8}$ and weak lensing data \cite{Lee:2012nh,Nesseris:2017vor,Kazantzidis:2018rnb,Gannouji:2018ncm,Gannouji:2020ylf}. It indicates a potential solution to resolve discrepancies in cosmological observations.
 
\begin{figure*}
\centering
\vspace{1cm}
\begin{tabular}{cc}
\includegraphics[width=0.5\linewidth]{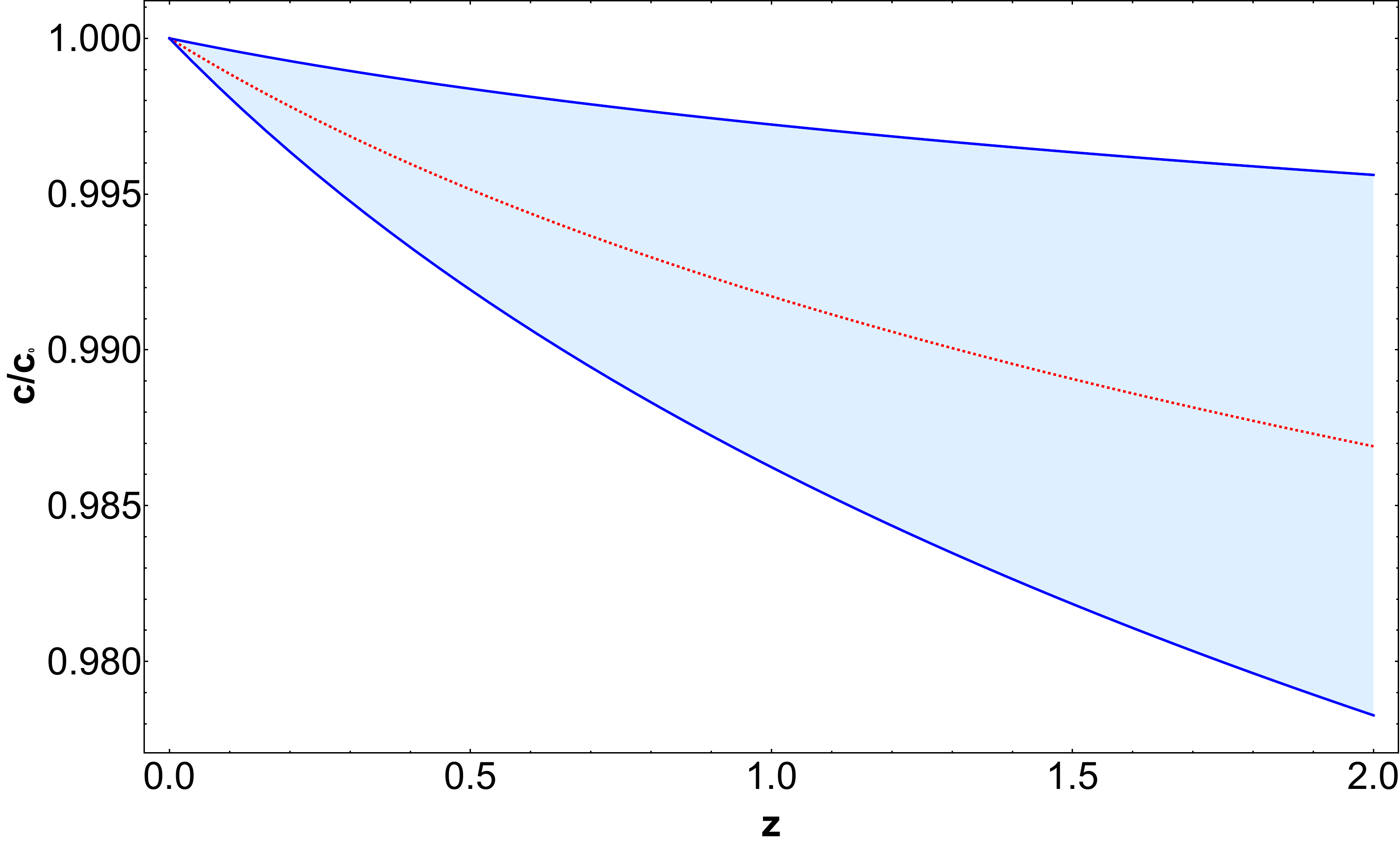} &
\includegraphics[width=0.5\linewidth]{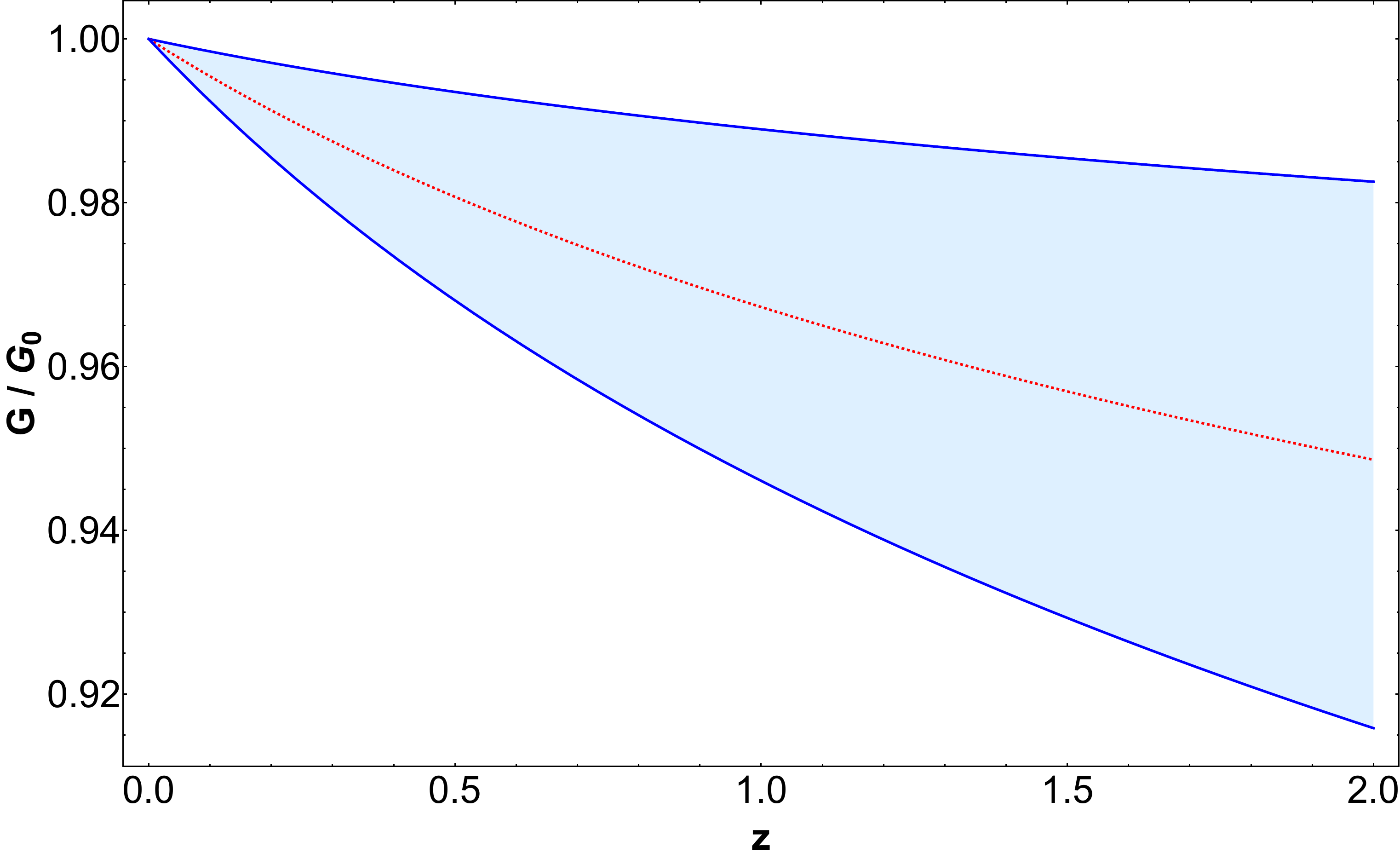}
\end{tabular}
\vspace{-0.5cm}
\caption{ It illustrates the evolution of the speed of light and Newton's gravitational constant over cosmic time, normalized to their current values. Left) The evolution of the speed of light, normalized to its current value, from $z=0$ to $z=2$. Right) The evolution of Newton's gravitational constant over the same redshift range.} \label{fig-1}
\vspace{1cm}
\end{figure*}
It indicates that the speed of light was between $0.4$\% and $2.2$\% slower at redshift $2$ compared to its current value. The left panel of Figure~\ref{fig-1} shows the evolution of the speed of light, normalized to its current value, over cosmic time. The central dashed line represents the best-fit values, while the upper and lower solid lines represent the $1$-$\sigma$ confidence levels. The right panel of Figure~\ref{fig-1} shows the evolution of the Newtonian gravitational constant, normalized to its current value, over cosmic time. Like the speed of light, the central dashed line represents the best-fit values, and the upper and lower solid lines represent the 1-$\sigma$ confidence levels for the gravitational constant. The gravitational constant could have decreased by $1.7$\% to $8.4$\% at $z=2$ compared to its current value, within the 1-$\sigma$ confidence levels.

\subsubsection{other bands without systematic error}
\label{subsubsec:otherbandnosys}

Table \ref{tab:Tab3} presents the upper bounds for the parameter $b$, which reflects potential variations in the speed of light and Newton's constant, obtained from different bands of data. The columns list the bands (\textit{e..g.}, $g$, $r$, $i$, $z$, all), the $b$-value ranges, the derived ratio of the speed of light at redshift $z=2$ to its current value $c_0$​, and the ratio of Newton's constant at redshift $z=2$ to its current value $G_0$. The final column, labeled SMC, indicates whether the results for each band are consistent with the SMC, where the speed of light and Newton’s constant are assumed to be constant. For the $g$, $r$, and $z$ bands, the data supports the SMC, as indicated by an O, while the i band shows inconsistency (marked with an X). The all row aggregates results across all bands, showing consistency with the SMC. The table suggests that the current DES data generally supports the constancy of these fundamental constants, except for the $i$ band.

 \begin{table}[h!]
 	\begin{center}
\begin{tabular}{|c|c|c|c|c|}
  \hline
  Bands & $b$-value & $c(z=2)/c_0$ & $G(z=2)/G_0$ & SMC \\
  \hline
  $g$ & $[-0.192\,,0.024]$ & $[0.993\,,1.054]$ & $[0.974\,,1.235]$ & O \\
  \hline
  $r$ & $[-0.048\,,0.016]$ & $[0.996\,,1.013]$ & $[0.983\,,1.054]$ & O \\
  \hline
  $i$ & $[0.016\,,0.080]$ & $[0.978\,,0.996]$ & $[0.916\,,0.982]$ & X \\
  \hline
  $z$ & $[-0.020\,,0.044]$ & $[0.988\,,1.006]$ & $[0.953\,,1.022]$ & O\\
  \hline
  all & $[-0.032\,,0.008]$ & $[0.998\,,1.009]$ & $[0.991\,,1.036]$ & O \\
  \hline
\end{tabular}
	\end{center}
\caption{These are the upper bounds for the range of $b$ obtained from each band and the possible variations in the speed of light and Newton's constant derived from them. The current DES data shows results consistent with the SMC (\textit{i.e.}, constant speed of light) for all bands except the $i$-band.}
\label{tab:Tab3}
\end{table}

\begin{itemize}
	\item $g$ band: The $b$-value can vary from $-0.192$ to $0.024$. This indicates possible deviations from the constant speed of light and Newton's constant. The ratio of the speed of light at $z=2$ to its current value, $c(z=2)/c_0$​, can vary from $0.993$ to $1.054$, meaning there is a small range for the speed of light to be slightly slower or faster at redshift $z=2$ compared to today. For Newton's constant, $G(z=2)/G_0$​, the values range from $0.974$ to $1.235$, suggesting slightly weaker or stronger.  The results are consistent with the SMC, as noted by the O.
	\item $r$ band: Here, the $b$-value ranges from $-0.048$ to $0.016$. The speed of light ratio can vary from $0.996$ to $1.013$, showing only minimal potential variation from its present value. Similarly, the ratio $G(z=2)/G_0$​ ranges from $0.983$ to $1.054$, indicating a limited possible deviation in Newton's constant. Like the $g$ band, the results are consistent with the SMC.
	\item $i$ band: The $b$-value for the i band ranges from $0.016$ to $0.080$. For the speed of light, $c(z=2)/c_0$​ could vary between $0.978$ and $0.996$, implying that the speed of light might have been slower at $z=2$. The ratio $G(z=2)/G_0$​ ranges from $0.916$ to $0.982$, suggesting that Newton's constant might also have been weaker.   The $i$ band is inconsistent with the SMC, as indicated by the X.
	\item $z$ band: The $b$-value in the $z$ band ranges from $-0.020$ to $0.044$. The speed of light ratio can vary between $0.988$ and $1.006$, meaning the speed of light could have been either marginally slower or faster at redshift $2$. The ratio $G(z=2)/G_0$​ ranges from $0.953$ to $1.022$, suggesting only minor variations in Newton's constant. The $z$ band is consistent with the SMC.
	\item All bands combined: When all bands are combined, the $b$-value ranges from $-0.032$ to $0.008$, providing more constrained upper limits for potential variations. The speed of light ratio​ lies between $0.998$ and $1.009$, while the ratio of Newton's constant​ varies from $0.991$ to $1.036$. This result suggests only slight variations, supporting the idea of constant fundamental constants. It indicates consistency with the SMC for the combined data across all bands.
\end{itemize}

\subsection{all-band with systematic error}
\label{subsec:allbandwsys}

\subsubsection{$i$-band with systematic error}
\label{subsubsec:ibandwsys}

A possible systematic effect due to the evolution of the stretch of the supernova light curve provides the value $n = 0.988 \pm 0.008 (\textrm{stat}) \pm 0.010 (\textrm{sys})$ including the maximum error. From this result, we obtain the value of the parameter $b$ of the meVSL model as
\begin{align}
b = 0.048 \pm 0.051 \label{bmeVSL2} \,.
\end{align}
From this value, the values of the speed of light and Newton's constant at $z=2$ are given as 
\begin{align}
0.973 \leq \frac{c(z=2)}{c_0} \leq 1.001 \quad , \quad
0.897 \leq \frac{G(z=2)}{G_0} \leq 1.003 \label{cGz2sys} \,, 
\end{align}
In this case, they are marginally consistent with the values in the standard model.
\begin{figure*}
\centering
\vspace{1cm}
\begin{tabular}{cc}
\includegraphics[width=0.5\linewidth]{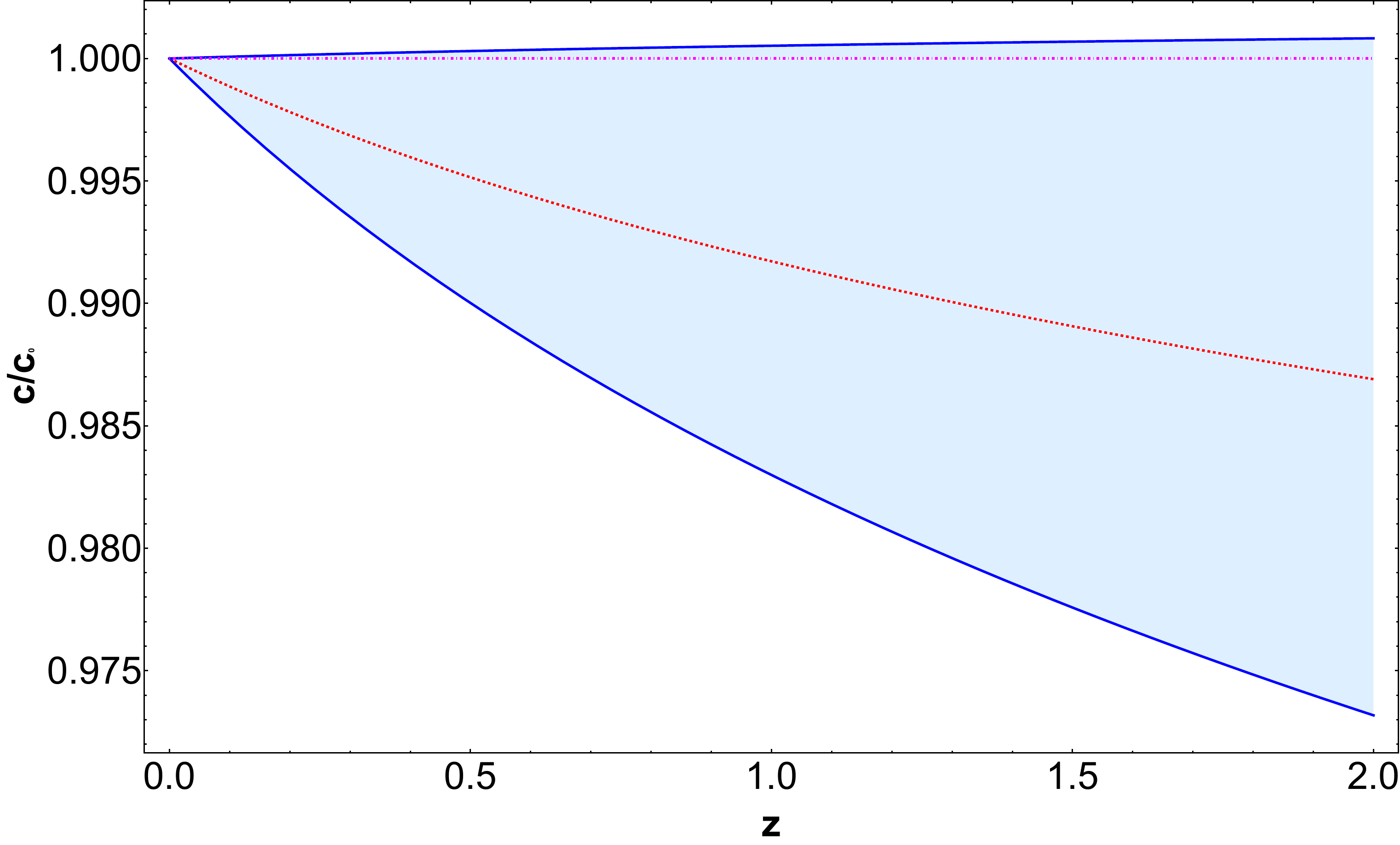} &
\includegraphics[width=0.5\linewidth]{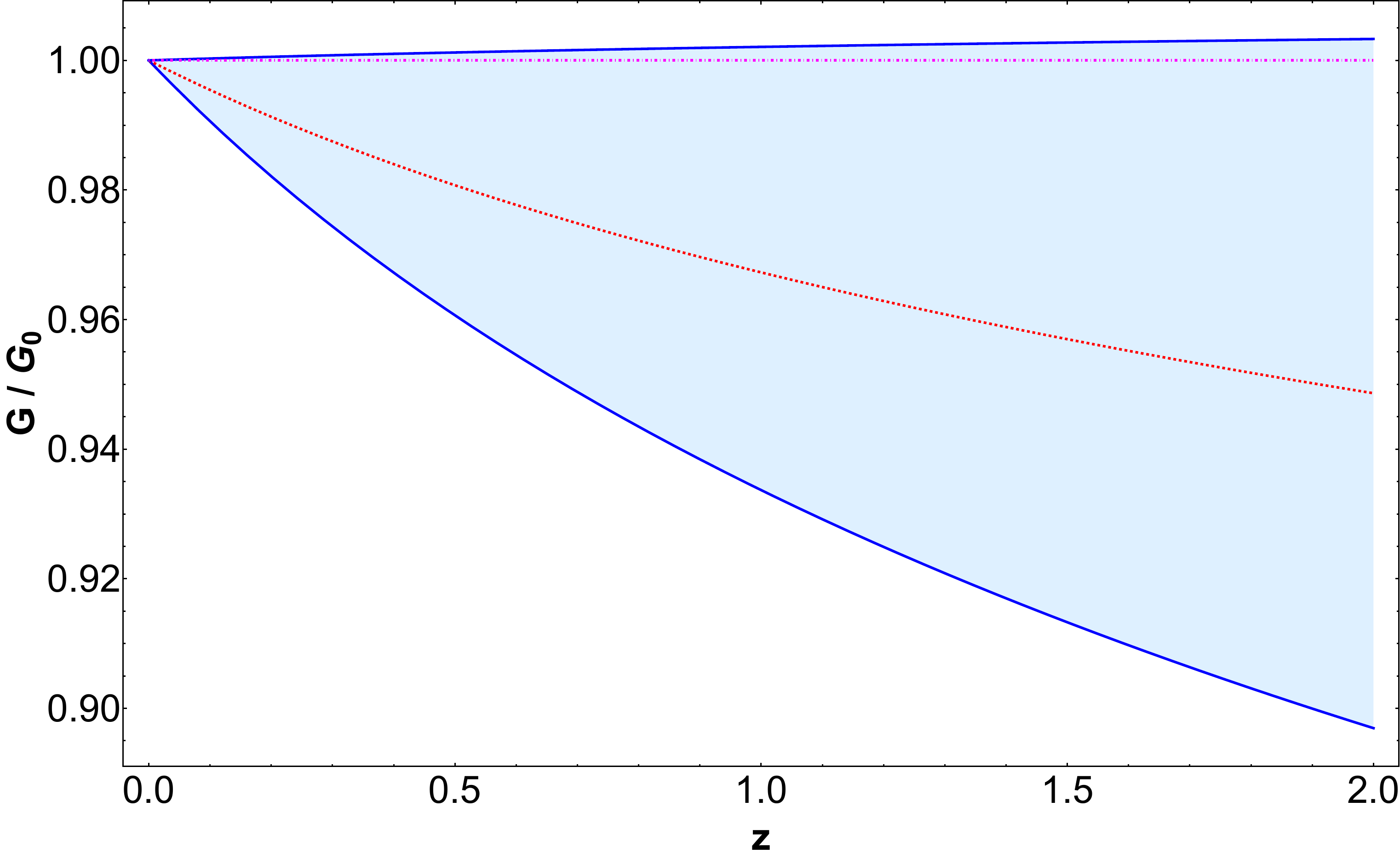}
\end{tabular}
\vspace{-0.5cm}
\caption{The possible evolutions of the speed of light and Newton's gravitational constant over cosmic time, normalized to their current values when we include the systematic error. Left) The evolution of the speed of light, normalized to its current value, from $z=0$ to $z=2$. Right) The evolution of Newton's gravitational constant over the same redshift range.} \label{fig-2}
\vspace{1cm}
\end{figure*}
 The left panel of Figure~\ref{fig-2} shows the evolution of the speed of light, normalized to its current value, over cosmic time. The central dashed line represents the best-fit values, the dot-dashed line denotes the null evolution, while the upper and lower solid lines represent the $1$-$\sigma$ confidence levels. The speed of light was up to $2.7$\% slower at redshift $2$ compared to its current value. The right panel of Figure~\ref{fig-2} shows the evolution of the Newtonian gravitational constant, normalized to its current value, over cosmic time. Like the speed of light, the central dashed line represents the best-fit values, and the upper and lower solid lines represent the 1-$\sigma$ confidence levels for the gravitational constant. The gravitational constant could have decreased to $10.4$\% at $z=2$ compared to its current value, within the 1-$\sigma$ confidence levels. However, both constants include values consistent with the standard model values.

\subsubsection{other bands with systematic error}
\label{subsubsec:allbandwsys}

 \begin{table}[h!]
 	\begin{center}
\begin{tabular}{|c|c|c|c|c|}
  \hline
  Bands & $b$-value & $c(z=2)/c_0$ & $G(z=2)/G_0$ & SMC \\
  \hline
  $g$ & $[-0.199\,,0.031]$ & $[0.992\,,1.056]$ & $[0.967\,,1.244]$ & O \\
  \hline
  $r$ & $[-0.067\,,0.035]$ & $[0.990\,,1.019]$ & $[0.962\,,1.076]$ & O \\
  \hline
  $i$ & $[-0.003\,,0.099]$ & $[0.973\,,1.001]$ & $[0.897\,,1.003]$ & O \\
  \hline
  $z$ & $[-0.039\,,0.063]$ & $[0.983\,,1.011]$ & $[0.933\,,1.044]$ & O\\
  \hline
  all & $[-0.057\,,0.033]$ & $[0.991\,,1.016]$ & $[0.964\,,1.065]$ & O \\
  \hline
\end{tabular}
\end{center}
\caption{These are the upper bounds for the range of b obtained from each band, and the possible variations in the speed of light and Newton's constant are derived from them when we include the systematic error. The current DES data shows results consistent with the standard model for all bands,  including the $i$-band.}
\label{tab:Tab4}
\end{table}

In Table \ref{tab:Tab4}, the results include systematic errors in the analysis of the parameter $b$, representing potential variations in the speed of light and Newton's constant at redshift $z = 2$. Here's the detailed explanation for each band:

\begin{itemize}
	\item $g$ band: The $b$-value varies from $-0.199$ to $0.031$, meaning there is a broader range of possible deviations from the standard model. The speed of light ratio $c(z=2)/c_0$ lies between $0.992$ and $1.056$, allowing for slight variations where the speed could be either slower or faster than its current value. Newton's constant $G(z=2)/G_0$ ranges from $0.967$ to $1.244$, suggesting possible deviations, but overall, the results are consistent with the SMC.
	\item $r$ band: In this band, the $b$-value ranges from $-0.067$ to $0.035$. The speed of light ratio lies between $0.990$ and $1.019$, indicating only minor deviations in the speed of light at redshift $z = 2$. The ratio of Newton's constant ranges from $0.962$ to $1.076$, suggesting potential variations in Newton's constant, but still consistent with the SMC.
	\item $i$ band: The $b$-value ranges from $-0.003$ to $0.099$, providing upper and lower bounds on possible deviations. For the speed of light, its ratio can vary from $0.973$ to $1.001$, meaning the speed of light could have been slightly slower at redshift $z = 2$. The ratio of Newton's constant ranges from $0.897$ to $1.003$, implying that Newton's constant could have been weaker. Despite the potential deviations, the i band supports the SMC when systematic errors are accounted for.
	\item $z$ band: The $b$-value in the $z$ band ranges from $-0.039$ to $0.063$. The speed of light ratio lies between $0.983$ and $1.011$. The ratio $G(z=2)/G_0$ ranges from 0.933 to 1.044, suggesting only minor variations in Newton's constant. The z band is consistent with the SMC.
	\item All bands combined: When data from all bands are combined, the $b$-value ranges from $-0.057$ to $0.033$, providing a more constrained upper bound for potential variations. The speed of light ratio lies between $0.991$ and $1.016$, indicating only slight possible deviations. The ratio of Newton's constant ranges from $0.964$ to $1.065$, showing limited potential variations in Newton's constant. The combined data supports the SMC.
\end{itemize}

In summary, after including systematic errors, all bands, including the $i$ band, show consistency with the SMC. Table \ref{tab:Tab4} demonstrates only slight potential variations in both the speed of light and Newton's constant, suggesting that the current DES data generally supports the idea of constant fundamental constants at redshift $z = 2$.

\subsection{Systematic from z}
\label{subsec:sysz}

We consider cosmological redshift using the Planck relation, which states the energy of a photon ($E$) is given by
\begin{align}
    E(a) = h \nu = h \frac{c}{\lambda} =
    \begin{cases}
        h_0 a^{-b/4} \frac{c_0 a^{b/4}}{\lambda_0 a} = h_0 \frac{c_0}{\lambda_0} a^{-1} & \text{meVSL} \\
        h_0 \frac{c_0}{\lambda_0 a} = h_0 \frac{c_0}{\lambda_0} a^{-1} & \text{SMC}
    \end{cases}
    = E_0 a^{-1} \label{Ea},
\end{align}
where $E_0 \equiv E(a=a_0=1)$. Therefore, by using the Planck relation for both SMC and meVSL, we obtain the same cosmological redshift relation. However, the Rydberg unit differs between the two models.  Using the cosmological evolution equations of the physical quantities described in Table~\ref{tab:LocalPhys}, we can derive the relationship of $E_{R}$ for $a$. In the meVSL model, the Rydberg energy scale $E_{R}$ is
\begin{align}
E_{R} = \frac{m_{e0} e_{0}^{4}}{2 \left( 4 \pi \epsilon_0 \right)^2 \hbar_0^2} a^{-\frac{b}{2}} \equiv E_{R0} a^{-\frac{b}{2}} \label{ERz} \,.
\end{align}
All atomic spectrum energy scales in the non-relativistic regime are described by the Rydberg unit $E_R$, and the cosmic evolution is incorporated into the calculation of the redshift parameter $z$, as expressed 
\begin{align}
\lambda_{i \textrm{emit}} &= \frac{hc}{E_R} \left( \frac{n_1^2 n_2^2}{n_2^2 - n_1^2} \right) = \frac{h_0c_0}{E_{R0}}  \left( \frac{n_{i}^2 n_{i+1}^2}{n_{i+1}^2 - n_{i}^2} \right) a_{\textrm{emit}}^{\frac{b}{2}} \equiv \lambda_{i \textrm{lab}} (1+z_{\textrm{emit}})^{-\frac{b}{2}} \nonumber \\ 
&= \lambda_{i 0} (1+z_{\textrm{emit}})^{-1}  \label{lambdainonrel} \,,
\end{align}
where $\lambda_{i \textrm{lab}}$ is the wavelength observed in the laboratory on Earth for a transition from level $i+1$ to level $i$, and $\lambda_{i 0}$ represents the wavelength of light emitted at $z_{\textrm{emit}}$ as observed on Earth. This introduces an additional factor in the measurement of $z$ in the meVSL model~\cite{Lee:2024part,Lee:2023xfg}
\begin{align}
z_{\textrm{emit}} = \left( \frac{\lambda_{i0}}{\lambda_{i\textrm{lab}}} \right)^{\frac{1}{1-b/2}} - 1 \label{zmeVSL} \,.
\end{align}
As can be seen from the above formula, when $b$ is positive (\textit{i.e.}, when the speed of light in the past is smaller than the current value and gradually increases), the value of $z_{\textrm{emit}}$ rises as the value of $b$ increases compared to the case where $b=0$ (i.e., when the speed of light is constant). Conversely, when $b$ is negative (i.e. when the speed of light in the past is larger than the current value and gradually decreases), the value of $z_{\textrm{emit}}$​ decreases as the value of $b$ decreases.

 \begin{table}[h!]
 	\begin{center}
\begin{tabular}{|c|c|c|c|c|c|}
  \hline
$b$ & $-0.05$ & $-0.01$ & $0$ & $0.01$ & $0.05$ \\
  \hline
 $z_{\textrm{emit}}$ & 0.966 & 0.993 & 1 & 1.007 & 1.036 \\
  \hline
$\Delta z_{\textrm{emit}}$ & -0.034 & -0.007 & 0 & 0.007 & 0.036 \\
  \hline
$|\Delta z_{\textrm{emit}}|/ z_{\textrm{emit}} (\%)$ & 3.5 & 0.69 & 0 & 0.70 & 3.59 \\
  \hline
\end{tabular}
\end{center}
\caption{These are changes in the redshift value obtained in the meVSL model depending on the value of $b$, compared to the known redshift value of $z=1$ in the SMC.}
\label{tab:Tab5}
\end{table}

Table \ref{tab:Tab5} presents how the redshift value ($z_{\textrm{emit}}$), its change ($\Delta z_{\textrm{emit}}$), and the percentage change ($|\Delta z_{\textrm{emit}}|/z_{\textrm{emit}}$​) are influenced by the parameter $b$. As the value of $b$ varies from $-0.05$ to $0.05$ in increments of $0.01$, the redshift value also changes, ranging from $0.966$ to $1.036$. The difference in redshift is calculated by comparing each redshift value to the baseline redshift value at $b=0$. This change fluctuates between $-0.034$ and $0.036$. The last row shows the percentage change in redshift relative to the original $z$​ value, with values ranging from $3.5$\% to $3.59$\%.

 \begin{table}[h!]
 	\begin{center}
\begin{tabular}{|c|c|c|c|c|c|}
  \hline
$b$ & $-0.05$ & $-0.01$ & $0$ & $0.01$ & $0.05$ \\
  \hline
 $z_{\textrm{emit}}$ & 1.921 & 1.984 & 2 & 2.017 & 2.086 \\
  \hline
$\Delta z_{\textrm{emit}}$ & -0.079 & -0.016 & 0 & 0.016 & 0.086 \\
  \hline
$|\Delta z_{\textrm{emit}}|/ z_{\textrm{emit}} (\%)$ & 3.97 & 0.82 & 0 & 0.83 & 4.31 \\
  \hline
\end{tabular}
\end{center}
\caption{These are changes in the redshift value obtained in the meVSL model depending on the value of $b$, compared to the known redshift value of $z=2$ in the SMC.}
\label{tab:Tab6}
\end{table}
We show how the redshift value ($z_{\textrm{emit}}$), its change ($\Delta z_{\textrm{emit}}$), and the percentage change ($|\Delta z_{\textrm{emit}}|/z_{\textrm{emit}}$​) vary with the parameter $b$ in Table \ref{tab:Tab6}. The values of $b$ range from $-0.05$ to $0.05$. For each value of $b$, the corresponding redshift changes, ranging from $1.921$ to $2.086$. The difference in redshift is calculated by comparing each redshift value to the baseline value when $b = 0$. The change in redshift fluctuates between $-0.079$ and $0.086$. We show the percentage of this change relative to the original redshift value, with the percentage ranging from $3.97$\% to $4.31$\% in the last row.

\section{Conclusion and Summary}\label{sec:Conc}

Using the recently released dark energy survey supernova program's $i$ band results, we applied the minimally extended varying speed of light model to investigate the cosmological evolution of the speed of light and Newton's constant within the 1-$\sigma$ confidence levels. Our analysis indicates that the speed of light was between $0.4$\% and $2.2$\% slower at redshift $2$ compared to its current value. Furthermore, the gravitational constant may have decreased by $1.7$\% to $8.4$\% at $z=2$ compared to its present value. However, this result is obtained when we ignore the systematic effect due to the evolution of the stretch of supernova light curves in the $i$-band results. If we include this systematic error, we can obtain results for the speed of light or Newton's constant that agree with the standard model even at the 1-sigma level.

These findings align with previous studies suggesting that Newton's constant $G(z)$ decreasing with redshift could resolve the observed tension between the Planck$15$ best-fit $\Lambda$CDM cosmological background and the $\Lambda$CDM background favored by growth $f\sigma_{8}$ and weak lensing data. However, the full-band cosmological time dilation data from the dark energy survey supernova program matches the results from the standard model cosmology, indicating that these findings alone are not yet conclusive evidence of a varying speed of light.  Additionally, high-z SNIa in the i-band may be affected by dust contamination or overwhelmed by noise, as indicated by slight deviations at wider widths.

It is important to note that the redshift $z$ values used in this data do not account for the additional effects predicted by the minimally extended varying speed of light model. For a consistent interpretation, the redshift values predicted by the minimally extended varying speed of light model should be used when analyzing the data. It necessitates recalculating the redshift for each dataset in the dark energy survey supernova program to compute the cosmological time dilation more accurately. 

For this reason, we cannot yet definitively conclude that the Dark Energy Survey Supernova Program data rigorously demonstrates any changes in the speed of light. Nevertheless, through this manuscript, we aim to emphasize that the DES SN program data serves as a critical resource for testing variations in the speed of light and other physical constants over cosmic time. Furthermore, with future public releases of data that include a more rigorous analysis of redshift evolution, we hope to confirm whether any reliable evidence for time variation in physical constants exists.

\section*{Acknowledgments}
SL is supported by the National Research Foundation of Korea (NRF), funded both by the Ministry of Science, ICT, and~Future Planning (Grant No. NRF-2019R1A6A1A10073079) and by the Ministry of Education (Grant No. NRF-RS202300243411).  


\end{document}